\documentclass[twocolumn,astrosymb,twocolappendix]{aastex7}
\usepackage{xspace}
\usepackage{CJKutf8}
\usepackage{bm}
\usepackage{appendix}
\usepackage{amsmath,amssymb}

\newcommand{\sbunit}{\mathrm{mag\ arcsec}^{-2}}

\newcommand{\elvesdwarf}{\textsc{ELVES-Dwarf}\xspace}

\newcommand{\code}[1]{\textbf{\texttt{#1}}}
\newcommand{\sersic}{S\'ersic\xspace}
\newcommand{\satgen}{\textsc{SatGen}\xspace}
\newcommand{\kms}{\mathrm{km\ s^{-1}}}

\newcommand{\nhosttng}{857\xspace}

\submitjournal{ApJL}

\usepackage{soul}

\shorttitle{Too-many-satellites Problem}
\shortauthors{Li et al.}

\graphicspath{{./}{figures/}}

\begin{document}
\begin{CJK*}{UTF8}{gbsn}

\title{A Possible ``Too-Many-Satellites'' Problem in the Isolated Dwarf Galaxy DDO~161}

\correspondingauthor{Jiaxuan Li}

\author[0000-0001-9592-4190]{Jiaxuan Li (李嘉轩)}
\email[show]{jiaxuanl@princeton.edu}
\affiliation{Department of Astrophysical Sciences, 4 Ivy Lane, Princeton University, Princeton, NJ 08540, USA}

\author[0000-0002-5612-3427]{Jenny E. Greene}
\email{jgreene@astro.princeton.edu}
\affiliation{Department of Astrophysical Sciences, 4 Ivy Lane, Princeton University, Princeton, NJ 08540, USA}

\author[0000-0002-1841-2252]{Shany Danieli}
\email{sdanieli@tauex.tau.ac.il}
\affiliation{Department of Astrophysical Sciences, 4 Ivy Lane, Princeton University, Princeton, NJ 08540, USA}
\affiliation{School of Physics and Astronomy, Tel Aviv University, Tel Aviv 69978, Israel}

\author[0000-0002-5382-2898]{Scott G. Carlsten}
\email{scarlsten@gmail.com}
\affiliation{Department of Astrophysical Sciences, 4 Ivy Lane, Princeton University, Princeton, NJ 08540, USA}

\author[0000-0002-7007-9725]{Marla Geha}
\email{marla.geha@yale.edu}
\affiliation{Department of Astronomy, Yale University, New Haven, CT 06520, USA}

\begin{abstract}
The abundance of satellite galaxies provides a direct test of $\Lambda$CDM and galaxy formation physics on small scales. While satellites of Milky Way-mass galaxies are well studied, those of dwarf galaxies remain largely unexplored. We present a systematic search for satellites around the isolated dwarf galaxy DDO~161 ($M_\star \approx 10^{8.4}\, M_\odot$) at a distance of 6 Mpc. We identify eight satellite candidates within the projected virial radius and confirm three new satellites through surface brightness fluctuation distance measurements from deep Magellan imaging data. Together with its confirmed satellite UGCA~319, DDO~161 has four confirmed satellites above $M_{\star}^{\rm sat} > 10^{5.4}\, M_\odot$, making it the most satellite-rich dwarf galaxy known to date. We compare this system with predictions from the TNG50 cosmological simulation, combined with currently established galaxy–halo connection models calibrated on Milky Way satellites, and find that DDO~161 has a satellite abundance far exceeding all current expectations. The rich satellite system of DDO~161 offers new insight into how low-mass galaxies occupy dark matter halos in low-density environments and may provide new constraints on the nature of dark matter.
\end{abstract}


\section{Introduction}\label{sec:intro}

The abundance of satellite galaxies provides a powerful test of both the $\Lambda$CDM cosmological model and the physics of galaxy formation on small scales. In $\Lambda$CDM, the hierarchical growth of structure predicts that dark matter halos of all masses should host dark matter subhalos, whose luminous counterparts are satellite galaxies \citep{Bullock2017}. Early comparisons between observations and theoretical predictions revealed a stark discrepancy: the Milky Way (MW) and M31 appeared to host far fewer satellites than the predicted number of dark matter subhalos, a tension known as the ``missing satellites problem'' \citep{Klypin1999,Moore1999}. Over the past two decades, this apparent conflict has been largely resolved by a combination of deeper, wide-field imaging surveys that revealed large populations of ultra-faint satellites \citep[e.g.,][]{Simon2019,Cerny2023,Savino2025,Tan2025} and theoretical advances showing that not every dark matter subhalo is able to form a galaxy. For instance, the ultraviolet background generated during cosmic reionization, which can strongly suppress gas accretion and cooling in low-mass subhalos, could prevent low-mass dark matter halos from forming a galaxy \citep[e.g.,][]{Bullock2000,Kravtsov2004,Sawala2016}. The number of satellites around MW–mass hosts is now broadly consistent with $\Lambda$CDM predictions once these effects are taken into account \citep{Kim2018,Nadler2020,Santos-Santos2022}. However, several recent studies have reported an excess of satellites around the Milky Way and other MW-like galaxies \citep[e.g.,][]{Homma2024,Muller2024,Taibi2025}, and similar trends have been noted for early-type hosts \citep{Kanehisa2024}, hinting at a possible ``too-many-satellites'' problem. We caution that some proposed satellites of these systems lack direct distance measurement and thus remain unconfirmed.

With the satellite systems of MW–like galaxies increasingly well characterized in the Local Volume and beyond \citep[e.g.,][]{CarlstenELVES2022,SAGA-III}, attention has recently turned to test an analogous prediction from the $\Lambda$CDM model: dwarf galaxies themselves, despite their lower masses, should also host their own satellites \citep[e.g.,][]{Dooley2017b,Dooley2017a,Sales2017}. While satellites of individual dwarf galaxies have been identified through targeted deep imaging studies \citep[e.g.,][]{Sand2015,Carlin2016,Carlin2024,Davis2021,Medoff2025,Doliva-Dolinsky2025}, systematic searches have only recently begun, most notably with the \elvesdwarf \citep {Li2025} and ID-MAGE \citep{Hunter2025} surveys. These surveys are assembling the first statistically meaningful samples of satellites of dwarf galaxies in the Local Volume. The number of confirmed satellites identified in these studies so far is broadly consistent with expectations from the $\Lambda$CDM framework and our current understanding of galaxy formation \citep{Li2025}, although the sample size remains small.

In this Letter, we present the satellite population of DDO~161, an isolated dwarf galaxy with a stellar mass of $M_\star \approx 10^{8.4}\ M_\odot$ at $D = 6$~Mpc and one of the hosts studied in the \elvesdwarf survey (\citealt{Li2025}; Li et al., in prep). Deep imaging and surface brightness fluctuation (SBF) distance measurements reveal a surprisingly rich satellite system with four satellites. The satellite stellar mass function significantly exceeds predictions from both cosmological simulations and semi-analytical models. This discovery indicates a possible ``too-many-satellites'' problem for dwarf hosts, and could offer new insights into galaxy formation in the low-mass and low-density regime. 

This paper is structured as follows. We describe the properties of DDO~161 and the satellite search in Section \ref{sec:detection}, then present distance measurements for the satellite candidates in Section \ref{sec:sbf}. In Section \ref{sec:results}, we show the satellite stellar mass function and satellite abundance for DDO~161, and compare with predictions from the cosmological simulation TNG50. In Section \ref{sec:discussion}, we check various factors that might alleviate the discrepancy between observation and theoretical predictions, and discuss the implications in Section \ref{sec:discussion}. In this work, we adopt a flat $\Lambda$CDM cosmology with the present-day matter density $\Omega_m = 0.3$ and a Hubble constant of $H_0 = 70\ \mathrm{km\, s^{-1}\, Mpc^{-1}}$. The photometric results are presented in the AB system \citep{Oke1983}. The stellar masses calculated in this work are based on a \citet{Kroupa2001} initial mass function. We apply a Milky Way dust extinction correction using the dust map in \citet{SFD1998} recalibrated by \citet{Schlafly2011}.

\begin{figure}
    \centering
    \includegraphics[width=1\linewidth]{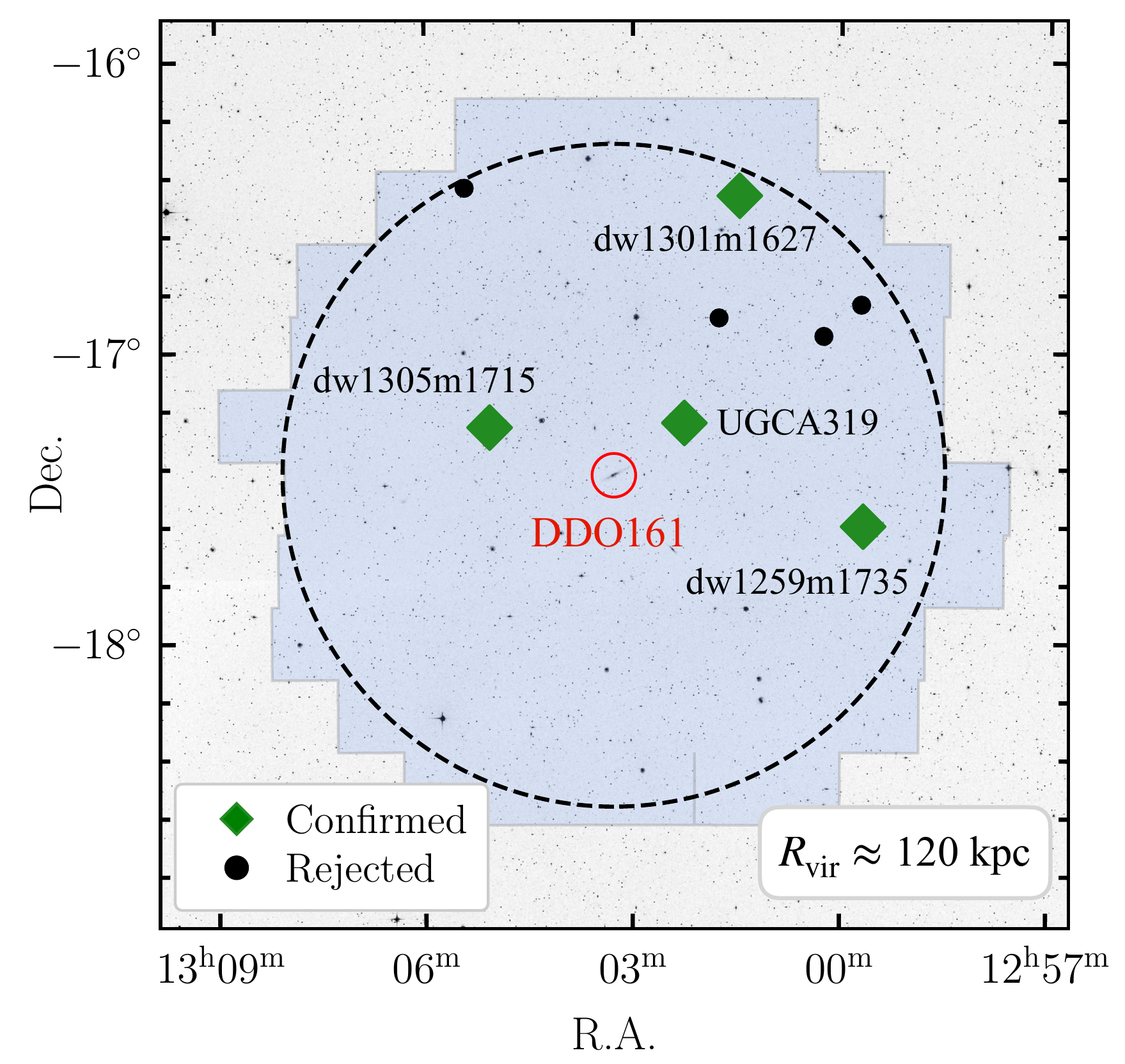}
    \caption{Satellite candidates of DDO~161. The footprint of the Legacy Surveys data used for satellite search is shown in blue. The black dashed circle corresponds to the projected virial radius of $R_{\rm vir} \approx 120\ \rm kpc$. The confirmed satellites are shown as green diamonds, and rejected candidates are shown as black dots. }
    \label{fig:coverage}
\end{figure}

\begin{figure*}
    \centering
    \includegraphics[width=0.83\linewidth]{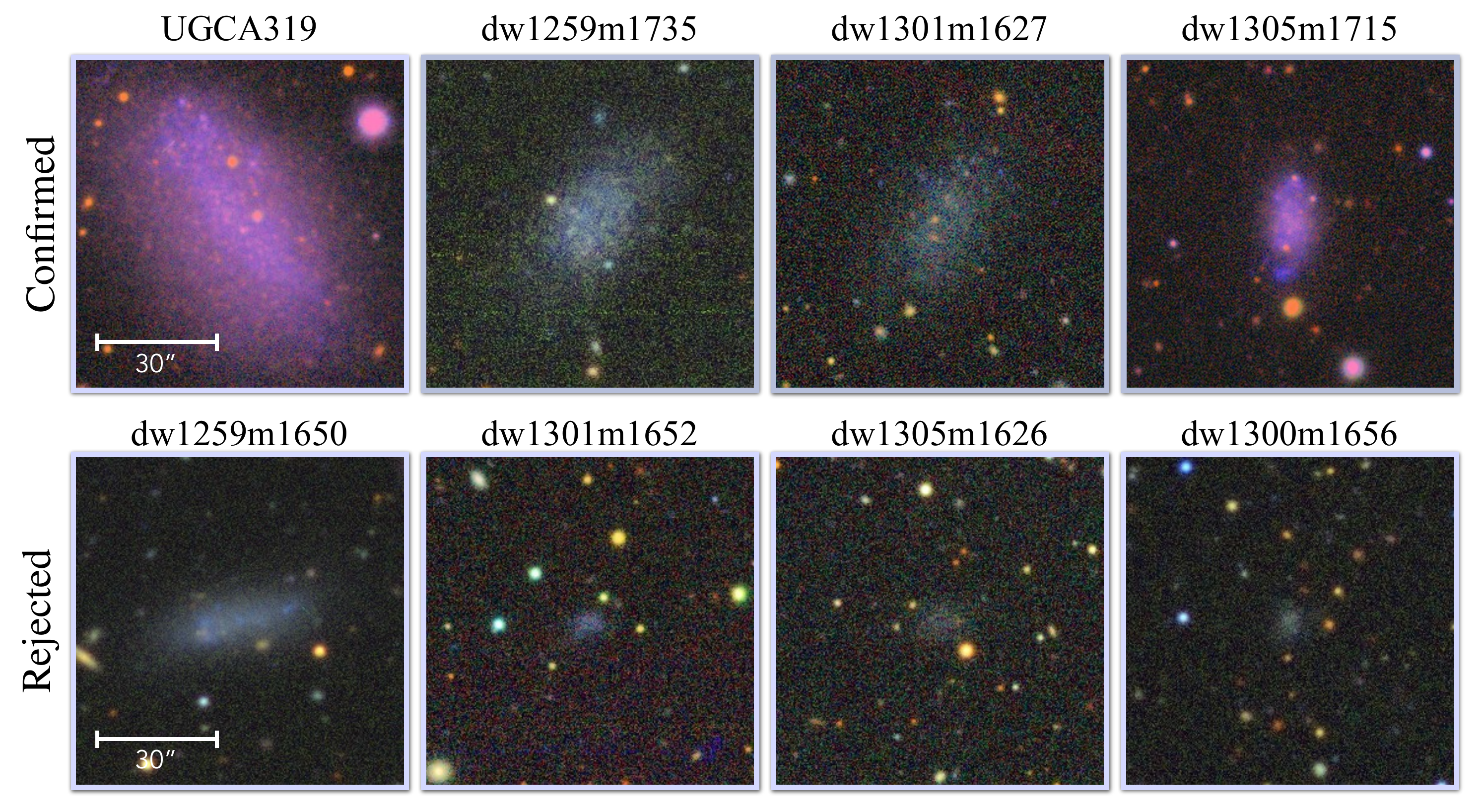}
    \caption{Cutout color-composite images of the satellite candidates of DDO~161 from the Legacy Surveys DR10. Cutouts are 80\arcsec{} on a side. The top row shows the four confirmed satellites, and the bottom row shows the four rejected candidates. UGCA~319 and dw1305m1715 only have $g$ and $i$-band data, lacking $r$-band coverage.}
    \label{fig:sat_images}
\end{figure*}

\section{Satellite Candidates around DDO161}\label{sec:detection}

\subsection{Properties of DDO~161}\label{sec:host}
DDO~161 is an isolated dwarf galaxy at a distance of $D_{\rm TRGB} = 6.03\pm 0.20$~Mpc, determined from the tip of the red giant branch (TRGB) method \citep{Karachentsev2017_DDO161}. Its heliocentric recession velocity is $v_h=744\ \mathrm{km\,s^{-1}}$ \citep{Meyer2004}. Prior to our investigation, DDO~161 was known to have one companion, UGCA~319, which lies 32 kpc away in projection and at a distance of $D_{\rm TRGB} = 5.75\pm 0.18$~Mpc \citep{Karachentsev2017_DDO161} with a similar heliocentric velocity of $v_h=750\ \mathrm{km\,s^{-1}}$ \citep{Kourkchi2020}, confirming their physical association. 

We estimate the stellar mass of DDO~161 as follows. It has apparent magnitudes of $B=13.50$~mag \citep{Makarov2014,Karachentsev2013} and $V=13.07$~mag \citep{Cook2014}, both in the Vega system. After correcting for Galactic and internal extinction following \citet{Karachentsev2013}, its absolute magnitude is $M_B=-16.04$~mag. Using the color--mass-to-light ratio ($M_\star/L$) relation from \citet{Bell2003}\footnote{$\log (M_\star/L)_B = 1.737 \cdot (B-V) - 1.092$}, we estimate its stellar mass to be $M_\star \approx 10^{8.3\pm 0.2}\, M_\odot$ based on the $B$-band photometry. \citet{Karachentsev2017_DDO161} reported a $K_s$-band luminosity of $\log L_{K_s}=8.75$, corresponding to $M_\star \approx 10^{8.5}\, M_\odot$ assuming $(M_\star/L)_{K_s}\approx 0.6$ \citep{Bell2003}. However, since this $K_s$ luminosity was not directly measured but inferred from $B$-band photometry, it should be used with caution. Other independent measurements yield $M_\star = 10^{8.11 \pm 0.38}\, M_\odot$ from \citet{Leroy2019} and $M_\star = 10^{7.91}\, M_\odot$ from the Wide-field Infrared Survey Explorer (WISE) mid-infrared photometry \citep{MHONGOOSE2024}. We checked the WISE data and found that its depth is not deep enough to capture the full extent of DDO~161’s star-forming disk, suggesting that the WISE-based stellar mass is likely underestimated.

Taking these results together, we adopt a fiducial stellar mass of $M_\star =10^{8.4}\,M_\odot$, and consider a range of $10^{8.1} < M_\star < 10^{8.7}\, M_\odot$ when comparing with theoretical predictions in \S \ref{sec:results}. This interval encompasses the published estimates of stellar mass from the literature, and places DDO~161 at a mass comparable to that of the Small Magellanic Cloud ($M_\star\approx10^{8.5}\, M_\odot$; \citealt{Skibba2012LMC}). Using the stellar-to-halo mass relation (SHMR) of \citet{RP2017}, we estimate a corresponding halo mass of $\log M_{\rm vir}/M_\odot = 11.0 \pm 0.15$, without including intrinsic scatter in the SHMR. The resulting virial radius, defined following \citet{Bryan1998} as the radius enclosing a mean density of $\Delta_{\rm vir} \approx 101.14$ times the critical density of the Universe, is $R_{\rm vir} \simeq 120 \pm 15$ kpc.


To understand the environment of DDO~161, we check the tidal indices from \citet{Karachentsev2013}, which quantify the local tidal field contributed by neighboring \textit{massive} galaxies. DDO~161 has a tidal index of $\Theta_1 = -0.9$, $\Theta_5=-0.5$, and $\Theta_j=-1.1$, all indicating that it resides in a weak external tidal field and is dynamically isolated from any massive neighbors. As pointed out by \citet{Karachentsev2017_DDO161}, DDO~161 is dynamically well separated from the nearest neighbor KK~176 ($\log M_\star/M_\odot \approx 7.44$). Its nearest massive galaxies are NGC~5068 ($\log M_\star/M_\odot \approx 9.73$, 1~Mpc away from DDO~161) and NGC~5236 ($\log M_\star/M_\odot \approx 10.86$, 1.8~Mpc away), confirming that DDO~161 is indeed very isolated. 

\subsection{Satellite Search}

\setlength{\tabcolsep}{2pt}
\begin{deluxetable*}{lCCcCCCCCCc}
\tablecaption{Photometric Properties of DDO~161 and Its Satellite Candidates}
\tablewidth{0pt}
\tablehead{
    \colhead{Name} & \colhead{R.A.} & \colhead{Decl.} & \colhead{$R_{\rm proj}$} & \colhead{$m_g$} & \colhead{$g-i$} & \colhead{$M_V$} & \colhead{$\log M_\star$} & \colhead{$r_{\rm eff}$} & \colhead{Comp.} & \colhead{Status} \\
    \colhead{} & \colhead{(deg)} & \colhead{(deg)} & \colhead{(kpc)} & \colhead{(mag)} & \colhead{(mag)} & \colhead{(mag)} & \colhead{($M_\odot$)} & \colhead{(kpc)} & \colhead{} & \colhead{}
}
\startdata
DDO~161 & 195.8184 & -17.4218 & 0 & \nodata & \nodata & \nodata & 8.40\pm 0.30 & \nodata & \nodata & Host \\
UGCA~319 & 195.5643 & -17.2416 & 32 & 14.22 \pm 0.08 & 0.35 \pm 0.04 & -14.82 \pm 0.08 & 7.50 \pm 0.07 & 0.81 \pm 0.04 & 1.00 & Confirmed \\
dw1259m1735 & 194.9186 & -17.5966 & 92 & 17.49 \pm 0.09 & 0.52 \pm 0.04 & -11.62 \pm 0.09 & 6.41 \pm 0.08 & 0.45 \pm 0.03 & 1.00 & Confirmed \\
dw1301m1627 & 195.3665 & -16.4598 & 110 & 17.89 \pm 0.10 & 0.63 \pm 0.04 & -11.26 \pm 0.10 & 6.39 \pm 0.08 & 0.50 \pm 0.03 & 1.00 & Confirmed \\
dw1305m1715 & 196.2662 & -17.2573 & 48 & 16.72 \pm 0.09 & 0.25 \pm 0.04 & -12.29 \pm 0.09 & 6.36 \pm 0.08 & 0.24 \pm 0.01 & 1.00 & Confirmed \\
dw1259m1650 & 194.9275 & -16.8348 & 109 & 17.32 \pm 0.09 & 0.46 \pm 0.04 & -11.77 \pm 0.09 & 6.40 \pm 0.08 & 0.45 \pm 0.03 & 1.00 & Rejected \\
dw1301m1652 & 195.4400 & -16.8801 & 69 & 19.77 \pm 0.13 & 0.33 \pm 0.06 & -9.27 \pm 0.13 & 5.24 \pm 0.11 & 0.15 \pm 0.02 & 0.90 & Rejected \\
dw1305m1626 & 196.3557 & -16.4339 & 117 & 20.02 \pm 0.14 & 0.84 \pm 0.06 & -9.13 \pm 0.14 & 5.81 \pm 0.12 & 0.21 \pm 0.02 & 1.00 & Rejected \\
dw1300m1656 & 195.0628 & -16.9428 & 91 & 20.83 \pm 0.17 & 0.73 \pm 0.09 & -8.34 \pm 0.18 & 5.34 \pm 0.16 & 0.12 \pm 0.02 & 0.96 & Rejected \\
\hline
\enddata
\tablecomments{Properties of DDO~161 and its satellite candidates, including name, R.A., Decl., projected distance from the host $R_{\rm proj}$, $g$-band apparent magnitude, $g-i$ color, $V$-band absolute magnitude ($M_V$), estimated stellar mass ($\log M_\star$), and half-light radius ($r_{\rm eff}$), completeness, and status (whether the candidate is confirmed as a satellite or rejected). For the satellite candidates, the stellar mass and half-light radius are estimated assuming the candidates are at the distance of the host, and are only meaningful for the confirmed ones. The estimated stellar mass of DDO~161 is described in \S \ref{sec:host}.
}
\vspace{-2em}
\label{tab:sats}
\end{deluxetable*}

We perform a systematic search for satellite candidates around DDO~161 using data from the Legacy Surveys Data Release 10\footnote{\url{https://www.legacysurvey.org/dr10/description/}} \citep{Dey2019}. We search for satellite candidates around DDO~161 within a projected radius of $\sim 120$~kpc. Our satellite detection closely follows \citet{CarlstenELVES2022} and \citet{Li2025}. First, we mask out bright stars by cross-matching with GAIA \citep{Gaia-dr3}. Then, we identify and replace bright sources and their associated diffuse light with sky noise by applying surface brightness thresholding. We smooth this cleaned image with a Gaussian kernel of 2.5--3.5 times the point-spread function (PSF) size to emphasize low surface brightness objects. Then we run Source Extractor \citep{SExtractor} in both the $g$ and $i$ bands with a low detection threshold of 2--2.5$\sigma$ to detect objects, and we only keep objects detected in both bands. All candidates were visually inspected to reject spurious detections, including scattered light from bright stars, galaxy outskirts, tidal features, and Galactic cirrus. 

For each remaining object, we fit a \sersic model to obtain the photometric and structural properties using \code{imfit} \citep{imfit}. Physical properties, such as stellar mass and physical size, are estimated assuming all objects lie at the host distance of 6 Mpc and adopting the $g-i$ color–$M_\star/L$ relation from \citet{Into2013}, which has been widely used in recent dwarf galaxy studies. We find consistent stellar masses when applying the newer calibration of \citet{delosReyes2025}. However, for consistency with our previous works \citep[e.g.,][]{CarlstenELVES2022,Li2025} and because $r$-band photometry is unavailable for two satellites, we adopt \citet{Into2013} in this work. Photometric uncertainties are estimated following \citet{ELVES-I} by injecting and recovering mock galaxies in the Legacy Surveys data. To further exclude background galaxies, we remove an additional 25 objects that are $>2\sigma$ outliers from the average mass--size relation of dwarf galaxies from \citet{ELVES-I}. As demonstrated in \citet{Li2025}, this criterion efficiently removes background systems whose inferred physical properties are inconsistent with those of dwarf galaxies at the assumed distance. For a population that intrinsically follows the mass–size relation, this cut would exclude only $\sim5\%$ of true satellites, corresponding to a negligible number given the small satellite abundance around dwarf hosts. Finally, we perform an additional round of visual inspection and remove six objects that exhibit clear morphological features characteristic of background galaxies, such as spiral arms or prominent bulge components.

In total, we identified eight satellite candidates within $R_{\rm vir} = 120$~kpc from DDO~161 in projection. Their basic properties are listed in Table \ref{tab:sats}, and their spatial distribution on the sky is shown in Figure \ref{fig:coverage}. The blue footprint indicates the coverage of the Legacy Surveys data used for the search. The cutout images of all candidates are shown in Figure \ref{fig:sat_images}. Three of these candidates (dw1305m1715, dw1259m1735, and dw1301m1627) were previously reported by \citet{Karachentsev2025}, who also suggested that Dw1309-1721 may be associated with DDO~161; we do not include this object as it lies outside our search footprint. We also note that dw1305m1715 has a heliocentric velocity of 725~$\kms$ based on the WALLABY survey data \citep{Mould2024}.

We quantify our search completeness by injecting mock galaxies into the Legacy Surveys data, as detailed in Appendix \ref{sec:completeness}. Our search is $>50\%$ complete at $M_g < -8$~mag and approaches unity at $M_g < -8.5$~mag for satellites lying within $1\sigma$ from the average mass--size relation. We therefore adopt $M_g=-8.5$~mag as our effective detection limit, corresponding to a stellar mass of $M_\star\approx 10^{5.4}\, M_\odot$ assuming an average color of $g-i=0.5$~mag. This limit is comparable to that in the \elvesdwarf survey \citep{Li2025} and is deeper than the average limit of the ELVES survey, due to our use of a larger smoothing scale and a lower detection threshold.

\section{Distance Measurement}\label{sec:sbf}

\begin{figure*}
    \centering
    \includegraphics[width=1\linewidth]{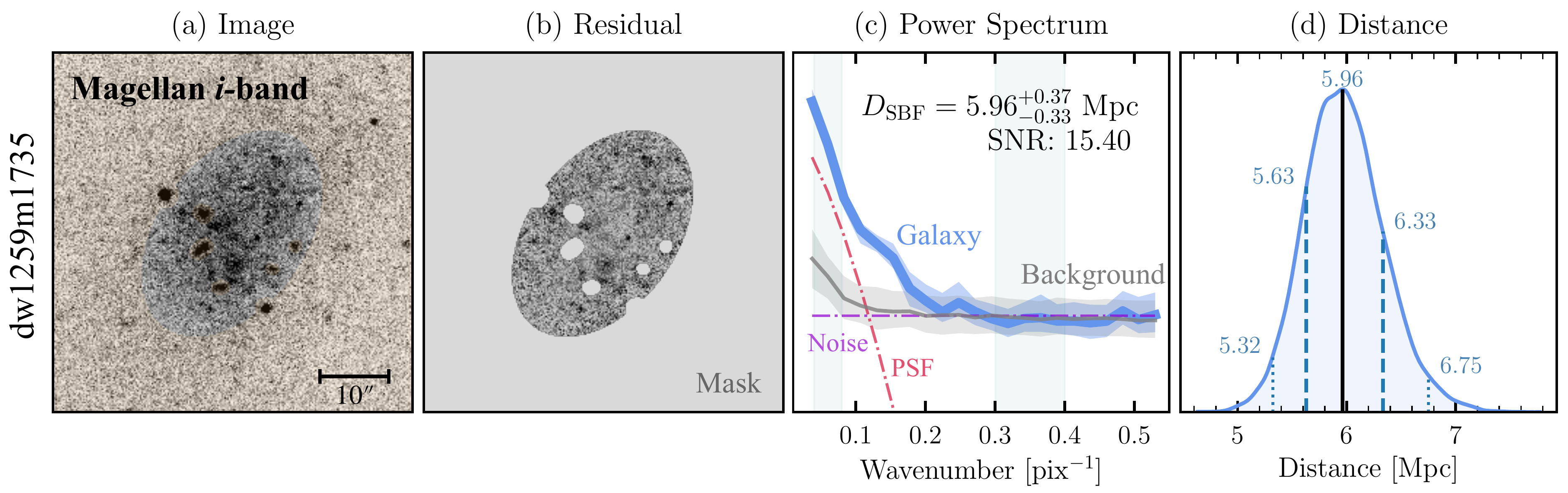}
    \vspace{-2em}
    \caption{Example SBF distance measurement, for a confirmed satellite dw1259m1735. (a) Magellan $i$-band image. (b) Residual image after subtracting a smooth galaxy model and masking bright sources, on which the SBF signal is measured. (c) Azimuthally averaged power spectrum (blue) fitted with the PSF (red) and white-noise (purple) components. The gray shaded region marks the contribution from the unmasked background sources. (d) SBF distance distribution showing the median, $1\sigma$ (16th–84th percentile), and $2\sigma$ (2.5th–97.5th percentile) intervals. The SBF distance agrees with the host distance of 6 Mpc, confirming dw1259m1735 as a satellite of DDO~161.}
    \label{fig:sbf_demo}
\end{figure*}

\begin{figure}
    \centering
    \includegraphics[width=1\linewidth]{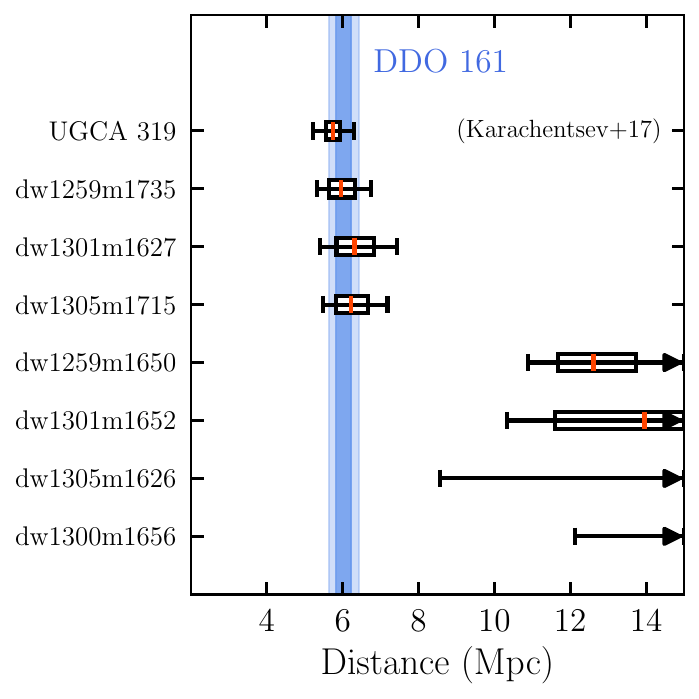}
    \caption{Distances of satellite candidates associated with DDO~161. The blue vertical line and shaded region indicate the TRGB distance of DDO~161 ($D=6.03\pm0.20$~Mpc). For each satellite, the red tick marks the median distance, while the box and whiskers show the $1\sigma$ (16th–84th percentile) and $2\sigma$ (2.5th–97.5th percentile) ranges of the measured distances, respectively. Arrows correspond to distance lower limits. For UGCA~319, the TRGB distance from \citet{Karachentsev2017_DDO161} is adopted, whereas all other distances are from SBF measurements in this work.}
    \label{fig:dist}
\end{figure}

\begin{deluxetable*}{lCCCc}\label{tab:dist}
\tabletypesize{\small}
\tablecaption{SBF Distances of DDO~161 and its Satellite Candidates}
\tablewidth{0pt}
\tablehead{
    \colhead{Name} & \colhead{$D_{\rm SBF}$} & \colhead{(S/N)$_{\rm SBF}$} & \colhead{$v_{h}$} & \colhead{Status} \\
    \colhead{} & \colhead{(Mpc)} & \colhead{} & \colhead{(km~s$^{-1}$)} & \colhead{}
}
\startdata
DDO~161 & \nodata & \nodata & 744 &  Confirmed \\
UGCA~319 & \nodata & \nodata & 750 & Confirmed \\
dw1305m1715 & $6.21^{+0.46,0.97}_{-0.39,0.73}$ & 20.2 & 725 & Confirmed \\
dw1259m1735 & $5.96^{+0.37,0.79}_{-0.33,0.64}$ & 15.4 & \nodata & Confirmed \\
dw1301m1627 & $6.31^{+0.52,1.13}_{-0.47,0.91}$ & 8.9 & \nodata & Confirmed \\
dw1259m1650 & $12.61^{+1.11,2.44}_{-0.93,1.73}$ & 8.2 & \nodata & Rejected \\
dw1301m1652 & $13.96^{+4.78,\infty}_{-2.36,3.62}$ & 2.1 & \nodata & Rejected \\
dw1305m1626 & $>8.57$ & 0.7 & \nodata & Rejected \\
dw1300m1656 & $>12.12$ & -0.6 & \nodata & Rejected \\
\hline
\enddata
\tablecomments{SBF distances of satellite candidates, including name, median distance, and 1$\sigma$ and 2$\sigma$ uncertainties, the signal-to-noise ratio of SBF measurement, confirmation status, and heliocentric velocity if available. UGCA~319 has a TRGB distance of $5.75\pm 0.18$~Mpc from \citet{Karachentsev2017_DDO161}. For dw1305m1626 and dw1300m1656, the distances correspond to the $2\sigma$ distance lower limit. }
\vspace{-2em}
\end{deluxetable*}

Reliable distance measurements are essential to confirm that the satellite candidates identified in Section~\ref{sec:detection} are physically associated with DDO~161 rather than being foreground or background galaxies. Apart from UGCA~319, which has a published TRGB distance, none of the other candidates have existing distance or velocity measurements. Spectroscopic follow-up of such low-surface-brightness dwarfs is challenging, and TRGB distances require space-based imaging. As an alternative, we employ the surface brightness fluctuation (SBF) technique \citep{Tonry1988,Greco2021,Cantiello2023}, which enables distance measurements from deep ground-based imaging alone. While less precise than TRGB, the typical SBF distance uncertainty ($\sim 10\%$) is sufficient to distinguish true satellites at the host distance from background galaxies. The method has been extensively validated and successfully applied to diffuse dwarf galaxies in the ELVES and \elvesdwarf surveys \citep{CarlstenELVES2022,Li2025}.

To obtain the imaging data with sufficient resolution and depth required for SBF measurement, we carried out a follow-up campaign with the 6.5 m Magellan Telescope, targeting seven of the satellite candidates. UGCA~319 was not included because it already has a TRGB distance from \citet{Karachentsev2017_DDO161}. Each candidate was observed with the Inamori Magellan Areal Camera and Spectrograph \citep[IMACS;][]{Dressler2011} in the $i$ band, with a total integration time of 30--40 minutes per target under excellent seeing conditions (0\farcs5–0\farcs6). Data reduction followed the procedure described in \citet{Li2024}, including bias subtraction, flat-fielding, astrometric and photometric calibration, and stacking. Photometric calibration was based on the DECam Local Volume Exploration Survey \citep[DELVE;][]{DELVE2021,DELVE2022} DR2 catalog. The Magellan images of the seven satellite candidates are shown in Figure \ref{fig:imacs_images}. These data reach sufficient depth and resolution to measure SBF signals for galaxies at $\sim$6~Mpc.

We measure SBF distances following the procedure described in \citet{Carlsten2019} and \citet{Li2025}. The SBF technique exploits the fact that the pixel-to-pixel flux variance of a galaxy image depends on distance: as distance increases, the galaxy appears progressively smoother. These fluctuations arise from Poisson variations in the number of bright stars per resolution element (see \citealt{Cantiello2023} for a review). 

An example SBF measurement in this work is illustrated in Figure~\ref{fig:sbf_demo}. For each target, we first construct a smooth model of the galaxy by fitting a \sersic model to the image (panel a), then build the residual image by subtracting the smooth model and dividing by the square root of the smooth model (panel b). We also mask out compact sources such as globular clusters and unresolved background galaxies by excluding sources brighter than $M_i < -5.0$~mag. We then compute the azimuthally averaged power spectrum of the residual image, which is modeled as the sum of a PSF power spectrum and a constant white-noise floor (panel c). The amplitude of the PSF component represents the measured SBF signal. To correct for contamination from unmasked background sources, we repeat the same analysis on blank fields near each target and subtract their contribution. The measured SBF signal is converted to a distance (panel d) using the $i$-band SBF calibration in \citet{Carlsten2019}. Distance uncertainties are estimated via Monte Carlo sampling over the galaxy color uncertainty, power spectrum fitting range, masking threshold, and background field choice.

The resulting SBF distances for the seven candidates are listed in Table~\ref{tab:dist} and summarized in Figure~\ref{fig:dist}. We classify a satellite candidate as confirmed if its SBF distance agrees with the host distance within $2\sigma$ and the SBF signal-to-noise ratio (S/N) exceeds S/N$>5$. Candidates whose $2\sigma$ distance range does not overlap with the host distance are rejected. Based on these criteria, three satellites are confirmed, and four are rejected, as demonstrated in Figure \ref{fig:dist}. The confirmed systems are visibly semi-resolved even in the Legacy Surveys imaging. For UGCA~319 which we do not have Magellan imaging, we use the TRGB distance from \citet{Karachentsev2017_DDO161}. Its distance agrees with the host distance within $2\sigma$, thus we include it as a confirmed satellite. In total, DDO~161 has four confirmed satellites.

For the confirmed satellites, we calculate their absolute magnitudes, stellar masses, and physical sizes assuming the host distance of 6.03~Mpc (see Table~\ref{tab:sats}). Variations in distance across the host’s virial volume ($D\pm R_{\rm vir}$) change the inferred stellar masses by less than 0.03 dex, which is well below the photometric uncertainties. The quoted stellar mass uncertainties include contributions from both photometric errors and the distance variation within the virial volume, but do not account for systematic uncertainties in the adopted color--$M_\star/L$ relation.
The most massive satellite, UGCA~319, has a measured stellar mass of $\log M_\star/M_\odot = 7.50\pm0.07$, which agrees quite well with $\log M_\star/M_\odot=7.6$ from \citet{Leroy2019} and the $K_s$-luminosity-based stellar mass of $\log M_\star/M_\odot=7.6$ from \citet{Karachentsev2017_DDO161} when assuming $\log M_\star/L_{K_s} = -0.4$.

\section{Results}\label{sec:results}

\begin{figure*}
    \centering
    \includegraphics[width=1\linewidth]{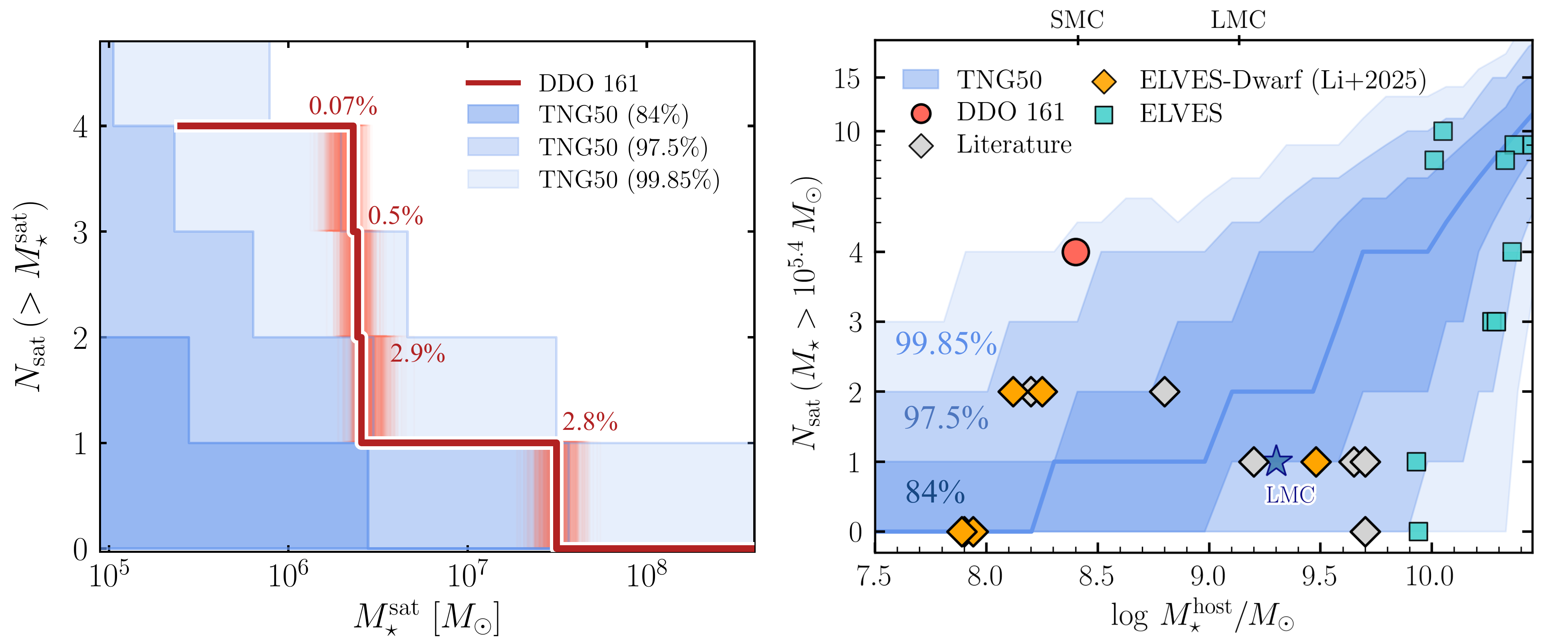}
    \vspace{-2em}
    \caption{\textit{Left}: cumulative satellite stellar mass function for DDO~161 (red). To account for the measurement uncertainties in stellar mass, we generate 1000 Monte Carlo realizations of the stellar mass function by resampling $M_\star^{\rm sat}$ within its uncertainty. Blue shaded regions show the 84th, 97.5th, and 99.85th percentiles of the TNG50 predictions, assuming the SHMR of \citet{Nadler2020}. Percentages marked along the red curve indicate the fraction of systems in TNG50 with an equal or greater number of satellites than observed. DDO 161 lies in the extreme tail of the distribution, hosting significantly more satellites than predicted. \textit{Right}: number of satellites per host above $M_\star^{\rm sat} > 10^{5.4}\, M_\odot$ as a function of host stellar mass. DDO~161 (red circle) is compared with hosts from individual literature studies (gray diamonds), the \elvesdwarf\ survey \citep[yellow diamonds;][]{Li2025}, and the ELVES survey \citep[turquoise squares;][]{CarlstenELVES2022}. Blue contours denote the 84th, 97.5th, and 99.85th percentiles of the TNG50 predictions. DDO~161 stands out clearly as an outlier, hosting a satellite population far richer than expected for its stellar mass.}
    \label{fig:mass_function}
\end{figure*}

We quantify the satellite population of DDO~161 in Figure \ref{fig:mass_function}. The left panel shows the cumulative satellite stellar mass function. To account for the measurement uncertainties in stellar mass, we generate 1000 Monte Carlo realizations of the stellar mass function by resampling each satellite’s measured $M_\star$ within its uncertainty, shown as the red shaded region. DDO~161 hosts one relatively massive companion, UGCA~319, with $M_\star \approx 10^{7.5}\, M_\odot$, roughly 1/10 of the host’s stellar mass. It has three additional satellites, all with $M_\star \approx 10^{6.4}\, M_\odot$, yielding a total of four confirmed satellites above our completeness limit of $M_\star^{\rm sat} > 10^{5.4}\, M_\odot$.

In the right panel of Figure~\ref{fig:mass_function}, we compare DDO~161’s satellite abundance with other dwarf hosts from the literature \citep[see Appendix E in][for a compilation]{Li2025}. Gray diamonds show systems within $D < 3.5$~Mpc that have been surveyed to at least a comparable depth, including the LMC, M33 \citep{Pace2025}, NGC~3109 \citep{Doliva-Dolinsky2025}, NGC~300 \citep{Sand2024}, NGC~55 \citep{Medoff2025}, NGC~2403 \citep{Carlin2024}, and NGC~4214 \citep{Carlin2021}. Among these, only NGC~3109 and NGC~55 are considered isolated. The yellow diamonds denote the eight isolated dwarf hosts from the first results of \elvesdwarf\ survey \citep{Li2025}, while turquoise squares show low-mass hosts from the ELVES survey \citep{CarlstenELVES2022}, only including confirmed satellites within the projected virial radius. DDO~161 stands out as a clear outlier, hosting more satellites than other dwarf hosts with comparable stellar mass, representing one of the richest satellite systems around a dwarf galaxy. Such a satellite abundance exceeds the model predictions from \citet{Sales2013} and \citet{Dooley2017b}.

To better interpret DDO~161's rich satellite population, we compare our observations with predictions from the TNG50 simulation \citep[][]{Pillepich2018,Pillepich2019,Nelson2019}, the highest-resolution simulation in the IllustrisTNG suite of cosmological hydrodynamical simulations. We use the TNG50-1 run, which follows the evolution of cosmic structure in a 50~Mpc box with a dark matter particle mass of $m_{\rm DM}=4.5\times10^5\, M_\odot$ and a baryonic mass resolution of $m_{\rm bar}=8.5\times10^4\, M_\odot$, making it capable of resolving satellite systems in dwarf-mass halos. Halos and subhalos are identified using the standard friends-of-friends (\textsc{FoF}) and \textsc{Subfind} algorithms \citep{Davis1985,Springel2001}. For each halo, we identify the central galaxy as the most massive subhalo, and classify all remaining subhalos as satellites.

With the mass resolution of TNG50, the central galaxies with stellar masses similar to DDO~161 are well resolved in both dark matter and stellar components. Therefore, we compute the stellar masses of central galaxies as the sum of all stellar particles within twice the stellar half-mass radius, following standard practice in TNG50 \citep[e.g.,][]{Shi2020,Engler2021}. In contrast, satellites with $M_\star\lesssim 10^7\,M_\odot$ are only marginally resolved \citep{Benavides2025}, although their dark matter subhalos remain well resolved. Therefore, we populate subhalos using the empirical SHMR.

To select systems analogous to DDO~161 in both stellar mass and environment, we first identify central galaxies with stellar masses in the range $10^{8.1} < M_\star^{\rm host} < 10^{8.7}\, M_\odot$, corresponding to DDO~161's estimated stellar mass range (\S\ref{sec:host}). We further require that no more massive halos reside within 1~Mpc to ensure an isolated environment. This results in a sample of \nhosttng analogs to DDO~161. The host halo masses have a median of $10^{10.85},M_\odot$, with a 2.5th--97.5th percentile range of $10^{10.5}$–$10^{11.2}\,M_\odot$, and the distribution is mildly skewed toward higher masses. For each satellite of the selected hosts, we obtain the peak dark matter mass ($M_{\rm peak}$) from its mass assembly history\footnote{For each host in TNG50, we adopt the \textsc{FoF}-group virial mass \texttt{Group\_M\_Crit200} from the TNG50 catalogs as host halo mass and \texttt{SubhaloMassInRadType[4]} as host stellar mass. For subhalos, we determine their peak dark matter mass by tracing \texttt{SubhaloMassType[1]} from the merger trees.}
and assign a stellar mass using the empirical SHMR of \citet{Nadler2020}, which is a popular galaxy-halo connection model calibrated on satellites of the Milky Way. To incorporate the intrinsic scatter in this relation, we generate 100 Monte Carlo realizations of the stellar masses for each satellite. In Section \ref{sec:discussion}, we also explore using alternative SHMRs and using the semi-analytical models, and get consistent results.

Two observational selection effects must be accounted for when comparing the observed satellite population with simulations. First, in observations, satellites are selected within the projected virial radius of the host on the sky (\S\ref{sec:detection}). To ensure a fair comparison, we apply an analogous projected selection to TNG50. For each host, we randomly project the satellite population onto a two-dimensional plane and select systems within the projected virial radius. This selection naturally includes a small number of objects that lie slightly beyond the 3D virial radius along the line of sight (e.g., recently accreted satellites or merging pairs) that would be included in observation. Compared with a strict 3D virial sphere cut, this selection increases the average satellite abundance by only $\Delta N_{\rm sat} = 0.07$. 

Second, although SBF measurements provide relatively precise distance estimates for the satellite candidates, the confirmed satellite sample may still include foreground or background galaxies that are scattered into the host distance due to SBF distance uncertainties. In this work, the $2\sigma$ SBF distance uncertainty is approximately 15\%, corresponding to $\sim0.9$~Mpc at a distance of 6~Mpc, which is substantially larger than the host virial radius ($\sim100$~kpc). To quantify the contamination fraction arising from SBF distance uncertainty, for each host in TNG50, we identify foreground and background galaxies that fall within the projected virial radius and would be classified as satellites given an assumed distance uncertainty of 15\% at a host distance of 6 Mpc and a stellar mass detection limit of $M_\star > 10^{5.4}\, M_\odot$. We define the contamination fraction as the ratio of such interlopers to the total number of galaxies that would be selected as satellites in observation. Applying this analysis to the DDO~161-like hosts in TNG50, we find a typical contamination fraction of $15\%$. This is consistent with previous estimates from \citet{Carlsten2021}, who reported a contamination fraction of $\sim 10-15\%$ for MW-mass hosts. Given the relatively small magnitude of this effect, we do not apply an explicit correction for contamination in this work. 

The predicted cumulative satellite stellar mass function from TNG50 is shown in the left panel of Figure~\ref{fig:mass_function}. The blue shaded regions mark the 84th, 97.5th, and 99.85th percentiles of the TNG50 prediction. The numbers marked along the red curve indicate the fraction of systems in TNG50 with more satellites than observed. Only 2.8\% of TNG50 hosts contain a massive satellite like UGCA~319, and 2.9\% have two satellites similar to DDO~161. When the third satellite is included, the fraction drops to 0.5\%. The probability for a DDO~161-like galaxy to host four satellites above $M_\star^{\rm sat}>10^{6.2}\, M_\odot$ is merely 0.07\%. Because the satellite count distribution is non-Gaussian \citep{Boylan-Kolchin2010,Mao2015}, we quantify the tension using percentiles rather than standard deviations. Nevertheless, the observed system lies at the 0.07\% tail of the TNG50 distribution, which would correspond to a deviation of $\sim3\sigma$ in the context of a Gaussian distribution.

The right panel of Figure~\ref{fig:mass_function} shows the predicted number of satellites per host above $M_\star^{\rm sat}>10^{5.4}\, M_\odot$ as a function of host stellar mass. The TNG50 predictions, shown as blue shaded regions, agree well with observations of other dwarf hosts from the literature, the \elvesdwarf\ survey, and the ELVES survey. DDO~161 still stands out as a pronounced outlier. The discrepancy is slightly smaller than that inferred from the satellite mass function because the satellite count is a less informative statistic than the full stellar mass function. 

To summarize, DDO~161 is the most satellite-rich dwarf galaxy found to date, and it is in clear tension with predictions based on the currently established galaxy-halo connection model calibrated on MW satellites.

\section{Discussion}\label{sec:discussion}

In this Letter, we present the discovery of an extremely rich satellite population of an isolated dwarf galaxy DDO~161. It has a stellar mass of $M_\star^{\rm host}\approx 10^{8.4}\, M_\odot$, similar to that of the Small Magellanic Cloud. DDO~161 hosts four satellites above $M_\star > 10^{5.4}\, M_\odot$ confirmed with direct distance measurements. This high satellite abundance represents an extreme outlier (at a 0.07\% level) compared with theoretical expectations derived from cosmological simulations and the empirical SHMR calibrated on Milky Way satellites \citep{Nadler2020}.

We estimate the likelihood of finding a system as extreme as DDO~161 within the Local Volume ($D<10$~Mpc; total volume about $4000~\mathrm{Mpc}^3$). In the TNG50 simulation, there are \nhosttng analogs to DDO~161 (in terms of both stellar mass and environment) within a $(50~\mathrm{Mpc})^3$ box, implying that only $\sim$30 such analogs should exist in the entire Local Volume ($<10$~Mpc). Among 30 systems, the chance of encountering an outlier at the 0.07\% level (Figure \ref{fig:mass_function}) is effectively zero. In the real Universe, the \citet{Karachentsev2013} catalog lists $\sim$200 dwarf galaxies with $10^{8}<M_\star<10^{9}\, M_\odot$ at $D<10$~Mpc, including both satellites and isolated systems. Among these galaxies, we would expect only $\sim$0.14 galaxies that are 0.07th-percentile outliers. Given that DDO~161 is one of only $\sim$30 isolated dwarf hosts targeted by the \elvesdwarf\ survey, the discovery of such a satellite-rich system is an exceedingly rare event, unlikely to be explained by statistical chance alone.

Observationally, reducing the tension between DDO~161 and theoretical predictions would require lowering the inferred stellar masses of its satellites by roughly 1 dex to bring the system into the 97.5th ($2\sigma$) percentile of the TNG50 distribution. As shown in the left panel of Figure \ref{fig:mass_function}, the measurement uncertainties alone (red shaded region) are not sufficient to alleviate the tension. The intrinsic scatter and systematic bias in the color–$M_\star/L$ relation are smaller than 0.2~dex \citep{delosReyes2025}. Moreover, stellar mass is often underestimated for a star-forming dwarf galaxy, as bright young stars can outshine the older population. Thus, uncertainties in the stellar mass estimates are likely insufficient to explain the discrepancy. It is also possible that DDO~161 resides in an unusually massive dark matter halo, which would increase the expected number of satellites; however, given existing stellar mass estimates and the steep slope of the SHMR at this mass scale, this effect is unlikely to fully reconcile the discrepancy in a statistical way.

\begin{figure}
    \centering
    \includegraphics[width=1\linewidth]{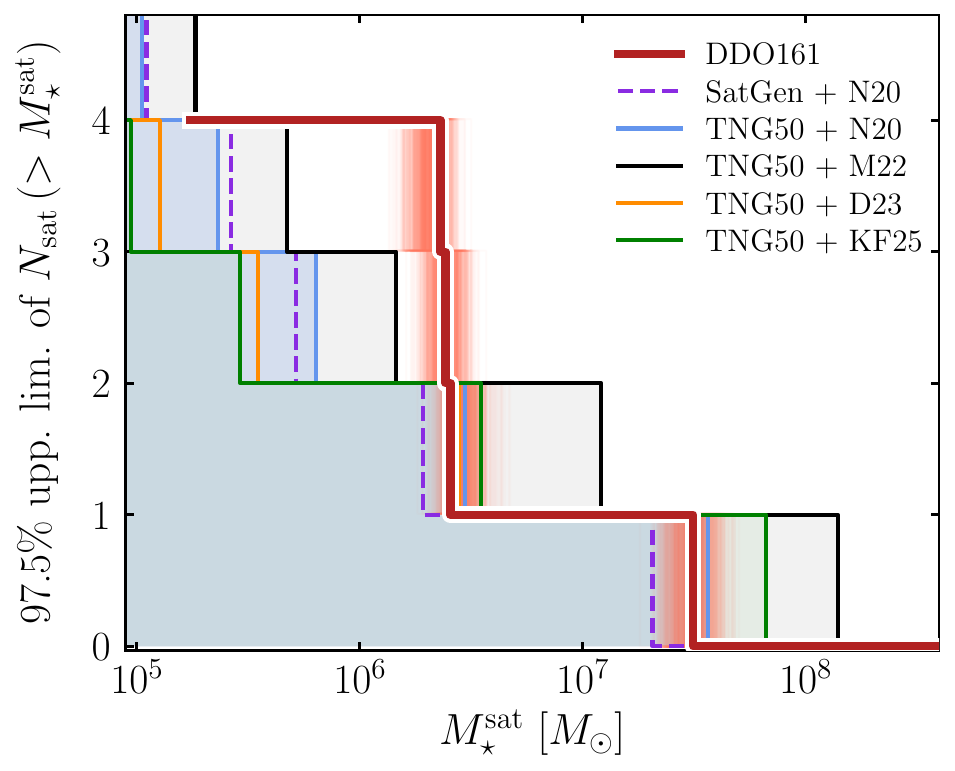}
    \caption{The predicted 97.5th percentiles of cumulative satellite stellar mass functions for DDO~161 analogs. The blue line shows results from the TNG50 simulation using the \citet{Nadler2020} SHMR, while the purple dashed line shows predictions from the semi-analytical model \satgen using the same SHMR. The close agreement between \satgen and TNG50 indicates that the discrepancy with DDO~161 is not driven by numerical effects in the simulation. 
    Additional curves show TNG50 predictions using alternative SHMRs from the literature (M22, \citealt{Manwadkar2022}; D23, \citealt{Danieli2023}; KF25, \citealt{Kado-Fong2025}). None of the currently established SHMRs can reproduce the high satellite abundance observed around DDO~161.}
    \label{fig:shmrs}
\end{figure}

We therefore consider possible explanations on the theoretical side. To test whether the discrepancy arises from the limitations of the TNG50 simulation, such as resolution and artificial subhalo disruption \citep{vdBosch2018,Benson2022}, we compare its predictions with those from the semi-analytical model \satgen\footnote{\url{https://github.com/JiangFangzhou/SatGen}} \citep{Jiang2021,Green2021,Monzon2024}. \satgen constructs halo merger trees and follows the evolution of subhalos under the influence of tidal stripping and heating from the host halo, dynamical friction, and baryonic effects. For this comparison, we take the host halo masses of the DDO~161 analogs identified in TNG50 and generate corresponding satellite populations using \satgen. Both TNG50 and \satgen therefore share the same host halo masses. 
For \satgen, the stellar mass of each satellite is then assigned using the same SHMR from \citet{Nadler2020}, based on subhalo peak mass. We note that \satgen only includes subhalos that are gravitationally bound within the 3D virial radius and therefore does not include satellites that lie just beyond the virial boundary. As discussed in \S\ref{sec:results}, the contribution from satellites beyond the 3D virial radius is negligible. In Figure~\ref{fig:shmrs}, the 97.5th ($2\sigma$) percentile line from \satgen (the purple dashed line) agrees closely with that from TNG50, confirming that the tension with DDO~161 is not likely driven by artifacts or resolution issues in TNG50. 

We then explore whether alternative SHMRs could help alleviate the tension, as different SHMRs essentially represent different galaxy formation efficiencies. In Figure~\ref{fig:shmrs}, we compare the 97.5th percentile of satellite stellar mass functions using several established SHMRs in the literature: \citet[][M22]{Manwadkar2022}, based on the semi-analytical model GRUMPY \citep{Kravtsov2022}; \citet[][D23]{Danieli2023}, calibrated on satellites of MW–like hosts from the ELVES survey; and \citet[][KF25]{Kado-Fong2025}, derived from the field dwarfs in the SAGA background sample \citep{Kado-Fong2024}. As shown in Figure~\ref{fig:shmrs}, DDO~161 remains an outlier across all models. The \citet{Manwadkar2022} relation yields the smallest discrepancy because it has a shallower slope and higher normalization than other SHMRs, but still with only $\sim$0.4\% of analog hosts exhibiting similarly rich satellite systems. None of the currently established SHMRs can reconcile the observed satellite abundance of DDO~161 with theoretical predictions.

A physical explanation for such an unusually satellite-rich system remains elusive. Most alternative dark matter models, including warm and self-interacting dark matter, predict \textit{fewer} low-mass subhalos than the cold dark matter paradigm \citep[e.g.,][]{Bullock2017,Gutcke2025}, and thus cannot explain the observed excess of satellites. Dynamically speaking, one possibility is that DDO~161 accreted UGCA~319 together with a bound group of companions, analogous to the Magellanic Clouds \citep{Sales2017}. However, such group infall events are already captured in cosmological simulations like TNG50 and therefore do not fully resolve the discrepancy. Another possibility is that DDO~161 resides within or near a filamentary structure, where recent infall or enhanced subhalo accretion could transiently elevate its satellite abundance. The TNG50 volume may not sample such environments adequately. Checking whether neighboring galaxies also show similarly enhanced satellite populations would help test this scenario.

A more plausible explanation is that the SHMR for satellites in dwarf groups differs from that calibrated for Milky Way–like systems, reflecting an environmental dependence \citep{Garrison-Kimmel2017,Read2017,Engler2021,Christensen2024}. However, other studies have found little or no environmental variation in the SHMR \citep[e.g.,][]{Watson2013,Shi2020,Danieli2023}. Simulations by \citet{Christensen2024} suggest a steeper SHMR in the field, whereas observations by \citet{Kado-Fong2025} find only weak environmental dependence in observation. For a given intrinsic scatter, a steeper SHMR would worsen the tension by producing even fewer satellites above a given stellar mass limit. On the other hand, a shallower relation, such as that proposed by \citet{Read2017} for field dwarfs in the Sloan Digital Sky Survey, could alleviate the discrepancy for DDO~161 but would overpredict satellites around other dwarf hosts. The real SHMR may also deviate from a single power law, following a piecewise or curved form that current models do not capture \citep{Santos-Santos2022}. In any case, DDO~161 provides powerful empirical constraints on the SHMR in the low-mass regime, particularly within dwarf groups where it remains poorly understood.

The discovery of such a satellite-rich dwarf galaxy underscores the importance of building a statistical sample of satellites around dwarf hosts. DDO~161 is part of the ongoing \elvesdwarf survey, which will provide a uniform census of $\sim$40 isolated dwarf hosts within 10~Mpc (J. Li et al., 2026, in preparation). If confirmed in additional systems, such ``too-many-satellites'' cases would offer powerful tests of $\Lambda$CDM on small scales and new insight into galaxy formation in the low-mass and low-density regime.

\appendix

\section{Satellite Detection Completeness}\label{sec:completeness}

\begin{figure}
    \centering
    \includegraphics[width=1\linewidth]{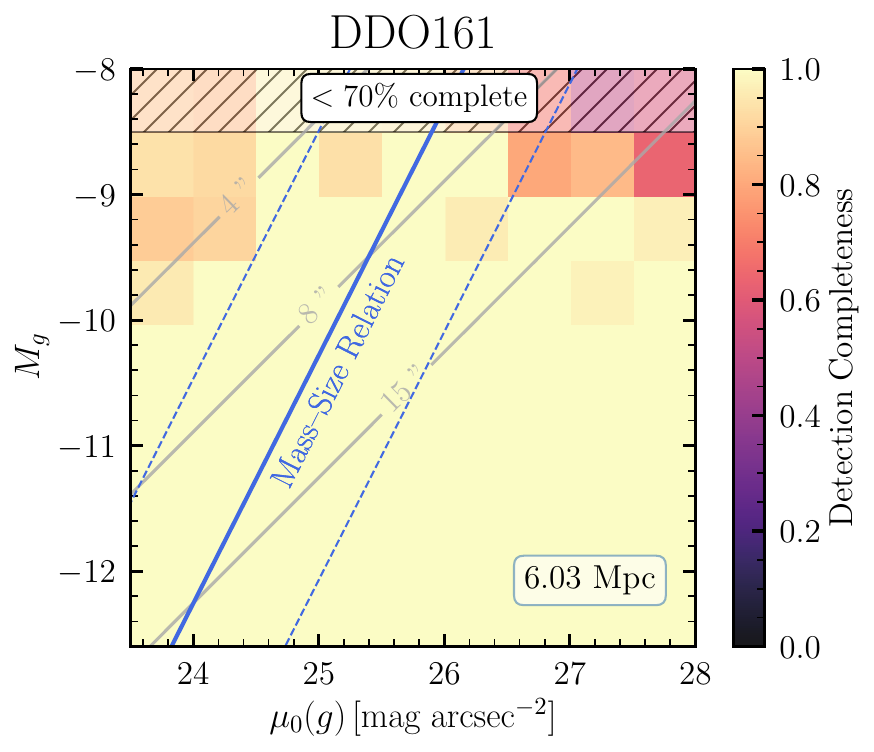}
    \caption{Detection completeness as a function of central surface brightness $\mu_0(g)$ and absolute magnitude $M_g$. The average mass--size relation of dwarf satellites from \citet{ELVES-I} is shown as the blue line. Our search is highly complete for satellites with $M_g < -8.5$~mag.}
    \label{fig:completeness}
\end{figure}

We quantify the completeness of our satellite search following the procedure described in \citet{Li2025}. Mock dwarf galaxies are injected into the same Legacy Surveys images used for the satellite detection. Each mock galaxy is modeled with a \sersic light profile with \sersic index $n=1$, and its properties are uniformly drawn in the following ranges: the absolute $g$-band magnitude in $-12.5 < M_g < -8$~mag, central surface brightness in $23.5 < \mu_0(g) < 28\ \sbunit$, color in $0.2<g-i<1.3$, and ellipticity in $0 < \varepsilon < 0.5$. Mock galaxies are injected into both $g$ and $i$-band images with a density of 12 per $0.25^{\circ}\times 0.25^{\circ}$ brick. Then the images went through the same detection pipeline. The detection completeness is thus defined as the ratio of recovered to injected galaxies. As shown in Figure \ref{fig:completeness}, we achieve a $>50\%$ completeness to $M_g\approx-8$~mag, and nearly unity for satellites brighter than $M_g \lesssim -8.5$~mag lying within $1\sigma$ of the mass--size relation.

\section{Magellan/IMACS Imaging of the Satellite Candidates}\label{sec:imacs_images}

\begin{figure*}
    \centering
    \includegraphics[width=0.83\linewidth]{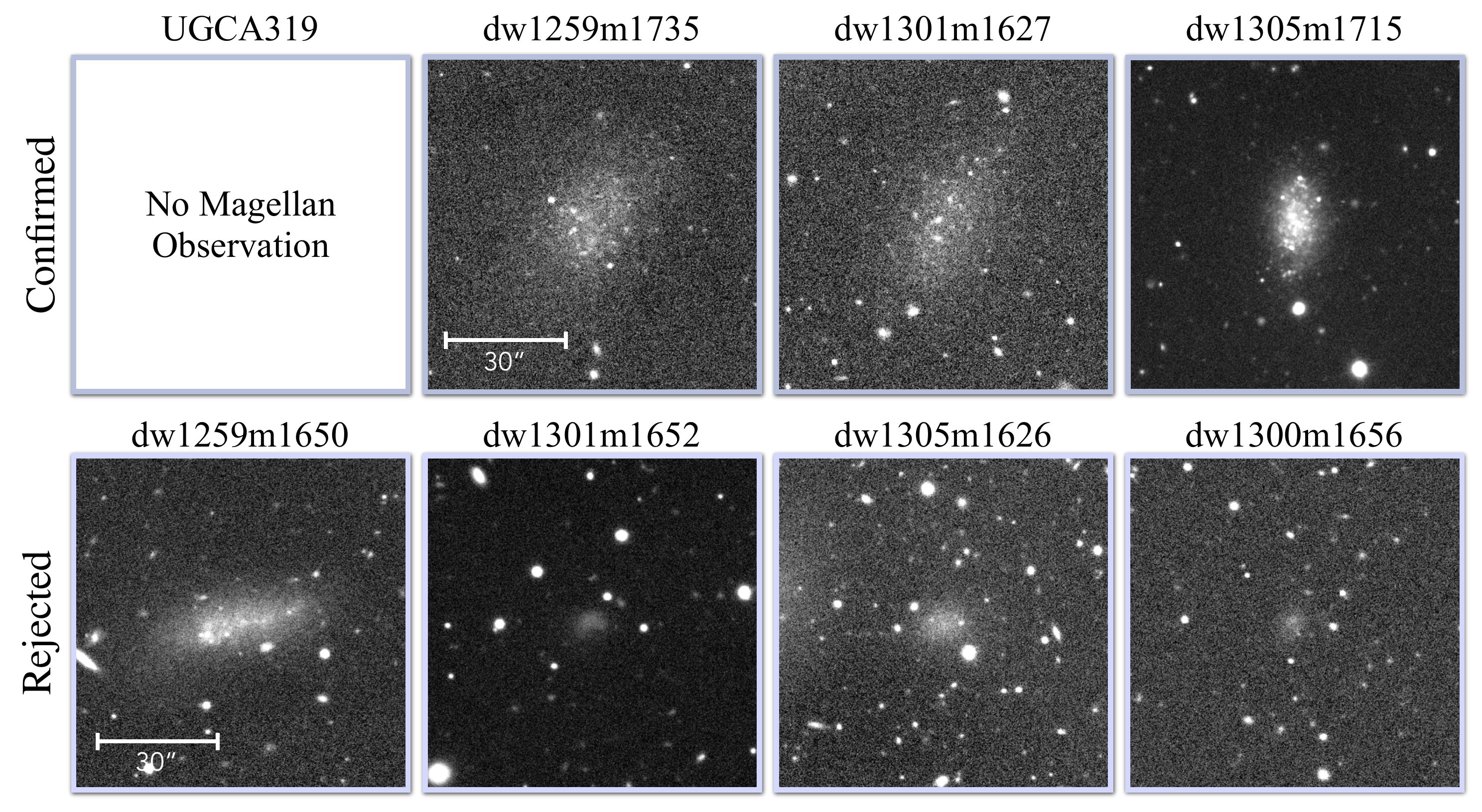}
    \caption{Magellan/IMACS $i$-band images of the seven satellite candidates of DDO~161. Cutouts are 80\arcsec{} on a side. UGCA~319 is not shown because no Magellan imaging was obtained for this galaxy. For comparison, the corresponding Legacy Surveys images are shown in Figure~\ref{fig:sat_images}.}
    \label{fig:imacs_images}
\end{figure*}

We present Magellan/IMACS $i$-band imaging for seven of the eight satellite candidates in Figure \ref{fig:imacs_images}. UGCA~319 is excluded because its distance is already established from a published TRGB measurement \citep{Karachentsev2017_DDO161}, and no additional Magellan imaging was obtained for this object. The three confirmed satellites (dw1259m1735, dw1301m1627, dw1305m1715; top row) all show prominent SBF signals, whereas the rejected objects (bottom row) have little or no SBF signal.

\section*{Acknowledgment}
J.L. is grateful for discussions with Andrey Kravtsov, Zhiwei Shao, Eric Bell, Alex Ji, Yao-Yuan Mao, Chin Yi Tan, Zhichao Zeng, Oleg Gnedin, Alyson Brooks, Sebastian Monzon, Anna Wright, and Tjitske Starkenburg. J.L. and J.E.G. gratefully acknowledge support from the NSF grant AST-2506292. 

This Letter includes data gathered with the 6.5-meter Magellan Telescopes located at Las Campanas Observatory, Chile. 

The Hyper Suprime-Cam (HSC) collaboration includes the astronomical communities of Japan and Taiwan, and Princeton University. The HSC instrumentation and software were developed by the National Astronomical Observatory of Japan (NAOJ), the Kavli Institute for the Physics and Mathematics of the Universe (Kavli IPMU), the University of Tokyo, the High Energy Accelerator Research Organization (KEK), the Academia Sinica Institute for Astronomy and Astrophysics in Taiwan (ASIAA), and Princeton University. Funding was contributed by the FIRST program from the Japanese Cabinet Office, the Ministry of Education, Culture, Sports, Science and Technology (MEXT), the Japan Society for the Promotion of Science (JSPS), Japan Science and Technology Agency (JST), the Toray Science Foundation, NAOJ, Kavli IPMU, KEK, ASIAA, and Princeton University. 

The Legacy Surveys consist of three individual and complementary projects: the Dark Energy Camera Legacy Survey (DECaLS; Proposal ID \#2014B-0404; PIs: David Schlegel and Arjun Dey), the Beijing-Arizona Sky Survey (BASS; NOAO Prop. ID \#2015A-0801; PIs: Zhou Xu and Xiaohui Fan), and the Mayall z-band Legacy Survey (MzLS; Prop. ID \#2016A-0453; PI: Arjun Dey). DECaLS, BASS and MzLS together include data obtained, respectively, at the Blanco telescope, Cerro Tololo Inter-American Observatory, NSF’s NOIRLab; the Bok telescope, Steward Observatory, University of Arizona; and the Mayall telescope, Kitt Peak National Observatory, NOIRLab. Pipeline processing and analyses of the data were supported by NOIRLab and the Lawrence Berkeley National Laboratory (LBNL). The Legacy Surveys project is honored to be permitted to conduct astronomical research on Iolkam Du’ag (Kitt Peak), a mountain with particular significance to the Tohono O’odham Nation. The Legacy Surveys imaging of the DESI footprint is supported by the Director, Office of Science, Office of High Energy Physics of the U.S. Department of Energy under Contract No. DE-AC02-05CH1123, by the National Energy Research Scientific Computing Center, a DOE Office of Science User Facility under the same contract; and by the U.S. National Science Foundation, Division of Astronomical Sciences under Contract No. AST-0950945 to NOAO.

NOIRLab is operated by the Association of Universities for Research in Astronomy (AURA) under a cooperative agreement with the National Science Foundation. LBNL is managed by the Regents of the University of California under contract to the U.S. Department of Energy.

This project used data obtained with the Dark Energy Camera (DECam), which was constructed by the Dark Energy Survey (DES) collaboration. Funding for the DES Projects has been provided by the U.S. Department of Energy, the U.S. National Science Foundation, the Ministry of Science and Education of Spain, the Science and Technology Facilities Council of the United Kingdom, the Higher Education Funding Council for England, the National Center for Supercomputing Applications at the University of Illinois at Urbana-Champaign, the Kavli Institute of Cosmological Physics at the University of Chicago, Center for Cosmology and Astro-Particle Physics at the Ohio State University, the Mitchell Institute for Fundamental Physics and Astronomy at Texas A\&M University, Financiadora de Estudos e Projetos, Fundacao Carlos Chagas Filho de Amparo, Financiadora de Estudos e Projetos, Fundacao Carlos Chagas Filho de Amparo a Pesquisa do Estado do Rio de Janeiro, Conselho Nacional de Desenvolvimento Cientifico e Tecnologico and the Ministerio da Ciencia, Tecnologia e Inovacao, the Deutsche Forschungsgemeinschaft and the Collaborating Institutions in the Dark Energy Survey. The Collaborating Institutions are Argonne National Laboratory, the University of California at Santa Cruz, the University of Cambridge, Centro de Investigaciones Energeticas, Medioambientales y Tecnologicas-Madrid, the University of Chicago, University College London, the DES-Brazil Consortium, the University of Edinburgh, the Eidgenossische Technische Hochschule (ETH) Zurich, Fermi National Accelerator Laboratory, the University of Illinois at Urbana-Champaign, the Institut de Ciencies de l’Espai (IEEC/CSIC), the Institut de Fisica d’Altes Energies, Lawrence Berkeley National Laboratory, the Ludwig Maximilians Universitat Munchen and the associated Excellence Cluster Universe, the University of Michigan, NSF’s NOIRLab, the University of Nottingham, the Ohio State University, the University of Pennsylvania, the University of Portsmouth, SLAC National Accelerator Laboratory, Stanford University, the University of Sussex, and Texas A\&M University.

BASS is a key project of the Telescope Access Program (TAP), which has been funded by the National Astronomical Observatories of China, the Chinese Academy of Sciences (the Strategic Priority Research Program “The Emergence of Cosmological Structures” Grant \# XDB09000000), and the Special Fund for Astronomy from the Ministry of Finance. The BASS is also supported by the External Cooperation Program of Chinese Academy of Sciences (Grant \# 114A11KYSB20160057), and Chinese National Natural Science Foundation (Grant \# 12120101003, \# 11433005).

The authors are pleased to acknowledge that the work reported in this paper was substantially performed using the Princeton Research Computing resources at Princeton University, a consortium of groups led by the Princeton Institute for Computational Science and Engineering (PICSciE) and the Office of Information Technology's Research Computing.

This research has made use of the SIMBAD database, operated at CDS, Strasbourg, France. The NASA/IPAC Extragalactic Database (NED) is funded by the National Aeronautics and Space Administration and operated by the California Institute of Technology. 

\vspace{1em}
\facilities{Blanco (DECam); Magellan:Baade (IMACS)}

\software{\href{http://www.numpy.org}{\code{NumPy}} \citep{Numpy},
          \href{https://www.astropy.org/}{\code{Astropy}} \citep{astropy:2013,astropy:2018,astropy:2022}, \href{https://www.scipy.org}{\code{SciPy}} \citep{scipy}, \href{https://matplotlib.org}{\code{Matplotlib}} \citep{matplotlib},
          \href{https://www.astromatic.net/software/sextractor/}{\code{SExtractor}} \citep{SExtractor},
          \href{https://www.astromatic.net/software/swarp/}{\code{SWarp}} \citep{swarp},
          \href{https://www.astromatic.net/software/psfex/}{\code{PSFEx}} \citep{psfex},
          \href{https://sep.readthedocs.io/en/v1.1.x/}{\code{sep}} \citep{Barbary2016},
          \href{https://www.mpe.mpg.de/~erwin/code/imfit/}{\code{imfit}} \citep{imfit},
          \href{https://github.com/kbarbary/sfdmap}{\code{sfdmap}},
          \href{https://github.com/JiangFangzhou/SatGen}{\code{SatGen} \citep{Jiang2021}}
          }

\bibliography{citation,software}
\bibliographystyle{aasjournal}

\newpage
\appendix

\end{CJK*}
\end{document}